\documentclass[12pt]{article}
\usepackage{amssymb,amsmath,amscd}
\newcommand{\lto}{\longrightarrow}

\newcommand{\E}{\mathcal{E}}

\newcommand{\Z}{\mathcal{Z}}
\newcommand{\rZ}{\text{Re }\mathcal{Z}}
\newcommand{\rrZ}{\text{Re }Z}
\newcommand{\iZ}{\text{Im }\mathcal{Z}}
\newcommand{\iiZ}{\text{Im }Z}
\newcommand{\F}{\mathcal{F}}

\newcommand{\sZ}{\mathsf{Z}}
\newcommand{\X}{\mathsf{X}}
\newcommand{\A}{\mathsf{A}}
\newcommand{\T}{\widehat{\mathbb{T}}^d}

\begin{document}
\begin{center}
{\Large \bf Mirror Duality and Noncommutative Tori}

\vspace{12mm} Eunsang Kim${}^{a,}$\footnote{eskim@hanyang.ac.kr}, \
and \; Hoil Kim${}^{b,}$\footnote{hikim@knu.ac.kr} \

\vspace{5mm} ${}^a${\it
Department of Applied Mathematics,\\ Hanyang University, Ansan Kyunggi-do 425-791, Korea}\\
${}^b${\it Department of Mathematics,\\
Kyungpook  National University, Taegu 702-701, Korea}\\

\vspace{12mm}

\end{center}

\begin{center}
 {\bf Abstract}

\vspace{5mm}

\parbox{125mm}{In this paper, we study a mirror duality on a
generalized complex torus and a noncommutative complex torus. First,
we derive a symplectic version of Riemann condition using mirror
duality on ordinary complex tori. Based on this we will find a
mirror correspondence on generalized complex tori and  generalize
the mirror duality on complex tori to the case of noncommutative
complex tori.

} \vfill
\end{center}
\setcounter{footnote}{0}

\pagebreak

\section{Introduction}
In this paper, we study  mirror dualities on complex tori,
generalized complex tori and noncommutative complex tori.

Based on string theory, it was proposed in \cite{SYZ} that mirror
pairs of Calabi-Yau manifolds admit special Lagrangian torus
fibrations over the same base such that generic fibers are dual
tori. In the case of abelian varieties there is a precise definition
of mirror duality which agrees with the suggestion of \cite{SYZ} via
dual torus fibratons, see \cite{Pol}, \cite{Fu1} and \cite{Tyu}.
Also, mirror symmetry on symplectic and complex tori or on abelian
varieties has been studied in many papers such as in \cite{Man,
GLO}. Based on the construction given in \cite{Pol, Fu1}, we
explicitly analyze the relations between Lagrangian submanifolds and
the holomorphic line bundles. From this we derive a symplectic
version of Riemann conditions and also find the mirror relation
between complexified symplectic form and period matrix for the dual
lattice. Furthermore, we find a condition for sumanifolds of a
symplectic torus which corresponds to a non-holomorphic line bundle
on the mirror dual complex torus.


The notion of generalized complex geometry, which was introduced by
N. Hitchin in \cite{Hit} and \cite{Gual}, contains as special cases
both complex and symplectic manifolds. It has been studied in
\cite{Gual} that in topological strings on a Calabi-yau manifold,
A-branes, B-branes and other corresponding notions discussed in
\cite{KaOr} are well explained in terms of generalized complex
submanifolds. In \cite{KaOr1}, Kapustin and Orlov studied the notion
of an $N=2$ superconformal vertex algebra and find a criterion for
two different complex tori to produce isomorphic $N=2$
superconformal vertex algebras which correspond to mirror duality.
In \cite{Kap}, it was discussed that the geometry of topological
D-branes is best described using generalized complex structures. In
particular, the role of B-field and the relation between T-duality
and an $N=2$ superconformal structure  are well explained in
\cite{Kap}. Based on the result in \cite{Kap} and the analysis made
on abelian varieties, we generalize the mirror duality on complex
and symplectic tori to the generalized complex tori. We shall
consider a special type of generalized complex torus and  define a
mirror map between generalized complex tori. Using this mirror map,
we verify the mirror correspondence on abelian varieties. Also, we
will discuss the case when a given B-field is not of type (1,1).

The noncommutative tori is known to be the most accessible examples
of noncommutative geometry developed by A. Connes, \cite{Connes94B}.
It also provides the best example in applications of noncommutative
geometry to string/M-theory which was initiated in \cite{CDS}.
Analogously the geometry and gauge theory of noncommutative torus
have been explicitly studied in many papers, such as \cite{CDS},
\cite{CoR}, \cite{Ri1}, \cite{Sch98}. A complex geometry of
noncommutative torus was developed by A. Schwarz in \cite{Sch01} and
it can be considered as a noncommutative generalization of abelian
varieties. It also provided a basic step to the study of M.
Kontsevich's homological mirror conjecture, \cite{Ko}. In
\cite{PolSch}, \cite{Kaj} and \cite{Pol1}, it has been shown that
the conjecture is true for the case of 2-dimensional noncommutative
tori. Based on the D-brane physics given in \cite{Kaj}, we discussed
the mathematical aspects of the T-duality on a noncommutative
complex torus in \cite{KK}. We generalize the mirror correspondence
to the higher dimensional cases. Also, we discuss the noncommutative
version of Riemann conditions.




\section{Mirror duality for Abelian varieties}

In this section, we briefly review the mirror symmetry on abelian
varieties following \cite{Fu1}, \cite{Pol}. We shall find a
necessary and sufficient condition that the mirror dual torus of a
symplectic torus becomes an abelian variety.

Let $\mathbb{T}^{d}=\mathbb{C}^g/(\mathbb{Z}\oplus i\mathbb{Z})^g$,
$d=2g$, be a complex torus equipped with a complexified symplectic
form $\Omega=\omega+i \xi$ such that
$\Omega\in\wedge^{1,1}(\mathbb{T}^{d})$. Let $V\cong \mathbb{R}^{d}$
be the universal cover of $\mathbb{T}^{d}$  and let
$\Gamma=\pi_1(\mathbb{T}^{d})\cong\mathbb{Z}^{d}$.  Since
$\Omega\in\wedge^{1,1}(\mathbb{T}^{d})$, there is an
$\Omega$-Lagrangian linear subspace $L$ of $V$ such that $L\cap
\Gamma\cong \mathbb{Z}^g$ and we can take $L=i\mathbb{R}^g\subset
\mathbb{C}^g$. Let us consider the Lagrangian torus fibration
$p:V/\Gamma\lto V/(L+\Gamma)$ which admits a section and hence we
have an isotropic decomposition $V=V/L\oplus L$ such that
$\Gamma=(\Gamma/\Gamma\cap L)\oplus(\Gamma\cap L)$. Let
$e_1,\cdots,e_g, e_{g+1},\cdots,e_{d}$  be a basis for the real
vector space $V=V/L\oplus L$, respectively, such that
$\Omega(e_i,e_j)=\Omega(e_{g+i},e_{g+j})=0$ and
$\Omega(e_{g+i},e_j)=\Z_{ij}$ for some $g\times g$ complex matrix
$\Z=(\Z_{ij})$. By the definition of $\Omega$, we may write
$\Z=\rZ+i\iZ$, where $\omega(e_{g+i},e_j)=(\rZ)_{ij}$ and
$\xi(e_{g+i},e_j)=(\iZ)_{ij}$. The symplectic form $\omega$ is
positive so that it is nondegenerate.  Note that the matrix $\Z$ can
be understood as a linear map from $V/L$ to $L^*$. In other words,
we define, using the same notation, $\Z:V/L\lto L^*$ by
$\Z(v)=\Omega(\cdot,v)$, $v\in V/L$. Similarly, we have linear maps
$\rZ$ and $\iZ$ such that $\rZ(v)=\omega(\cdot,v)$ and
$\iZ(v)=\xi(\cdot,v)$ for $v\in V/L$. Since  the matrix $\rZ$ is
nondegenerate,  the linear map $\rZ:V/L\lto L^*$ is an isomorphism.

Following the lines of \cite{Pol} and \cite{Fu1}, we construct the
mirror dual of $(\mathbb{T}^{d},\Omega)$. Note that the natural map
\[\alpha:V\oplus V^*\lto \text{Hom}_\mathbb{R}(V,\mathbb{C})\] defined by
$\alpha(v,v^*)(x)=\Omega(x,v)+iv^*(x)$, $x\in V$, is an isomorphism
of real vector spaces, where
$V^*=\text{Hom}_{\mathbb{R}}(V,\mathbb{R})$ is the dual vector space
of $V$. There exists a unique complex structure on $V\oplus  V^*$
induced by the isomorphism $\alpha$. Let $L^\perp=\{v^*\in V^*\mid
v^*(l)=0 \text{ for all } l\in L\}$. Then $\alpha$ maps the subspace
$L\oplus L^\perp\subset V\oplus V^*$ to the subspace
$\text{Hom}_\mathbb{R}(V/L,\mathbb{C})\subset
\text{Hom}_\mathbb{R}(V,\mathbb{C})$. Passing to the quotient
spaces, we get an isomorphism
\[\alpha_L:V/L\oplus L^*\lto
\text{Hom}_\mathbb{R}(L,\mathbb{C})=L^*\otimes_{\mathbb{R}}\mathbb{C}\]
where $L^*=\text{Hom}_{\mathbb{R}}(L,\mathbb{R})$. Indeed,
$\alpha_L$ is given as follows:
\begin{align*}
\alpha_L(v+L,l^*)(x)&=\Omega(x,v)+il^*(x)\\
&=\Omega(x,v)+i\omega(x,y),
\end{align*}
where we have used the isomorphism $\omega:V/L\lto L^*$ defined by
$\omega(y)=\omega(\cdot,y)$. Let us put
\[(\Gamma\cap L)^\perp=\{\mu\in L^*\mid \mu(\gamma)\in \mathbb{Z}
\text{ \ \ for all }\gamma\in\Gamma\cap L\}.\] Then the mirror of
$(\mathbb{T}^{d},\Omega)$ is defined to be
\[(\mathbb{T}^{d},\Omega)^\vee=W/\Lambda,\] where
\[W=(V/L)\oplus L^*, \ \ \ \Lambda=(\Gamma/\Gamma\cap L)\oplus (\Gamma\cap
L)^\perp\] A complex structure on $(\mathbb{T}^{d},\Omega)^\vee$ is
defined uniquely by the isomorphism $\alpha_L$. More explicitly, we
define a complex structure $\widehat{J}_\Omega$ on $V/L\oplus V/L$
which makes the following diagram commute:
\[\begin{CD} V/L\oplus
V/L@>\alpha_L>>\text{Hom}_{\mathbb{R}}(L,\mathbb{C})=L^*\otimes_{\mathbb{R}}\mathbb{C}\\
@V\hat{J}_\Omega VV  @VV \cdot iV\\
V/L\oplus
V/L@>\alpha_L>>\text{Hom}_{\mathbb{R}}(L,\mathbb{C})=L^*\otimes_{\mathbb{R}}\mathbb{C}
\end{CD}\]
Hence,
\begin{align}\nonumber
\widehat{J}_\Omega&=\begin{pmatrix} \text{Re
}\mathcal{Z}&0\\\text{Im }\mathcal{Z}&\text{Re
}\mathcal{Z}\end{pmatrix}^{-1}\begin{pmatrix}0&-1\\1&0\end{pmatrix}
\begin{pmatrix} \text{Re }\mathcal{Z}&0\\\text{Im
}\mathcal{Z}&\text{Re }\mathcal{Z}\end{pmatrix}\\
&=\begin{pmatrix} (\text{Re }\mathcal{Z})^{-1}&0\\-(\text{Re
}\mathcal{Z})^{-1}\text{Im }\mathcal{Z}(\text{Re
}\mathcal{Z})^{-1}&\text{Re
}\mathcal{Z}\end{pmatrix}\begin{pmatrix}0&-1\\1&0\end{pmatrix}
\begin{pmatrix} \text{Re }\mathcal{Z}&0\\\text{Im
}\mathcal{Z}&\text{Re }\mathcal{Z}\end{pmatrix}\nonumber\\
&=\begin{pmatrix} -(\text{Re }\mathcal{Z})^{-1}\text{Im
}\mathcal{Z}&-1\\1+(\text{Re }\mathcal{Z})^{-1}\text{Im
}\mathcal{Z}(\text{Re }\mathcal{Z})^{-1}\text{Im
}\mathcal{Z}&(\text{Re }\mathcal{Z})^{-1}\text{Im
}\mathcal{Z}\end{pmatrix}\label{cstru2}
\end{align}
For simplicity, let $\mathsf{A}=(\text{Re }\mathcal{Z})^{-1}\text{Im
}\mathcal{Z}$. Then we have
\begin{equation}\label{dcplex}
\widehat{J}_\Omega=\begin{pmatrix}-\mathsf{A}&-1\\1+\mathsf{A}^2&\mathsf{A}\end{pmatrix}.
\end{equation}
In fact, as a linear map, $\mathsf{A}:V/L\lto V/L$ is defined by the
condition
\[\omega(\cdot,\mathsf{A}v)=\xi(\cdot, v), \ \ v\in V/L.\]

Let $f:V/L\lto L$ be an $\mathbb{R}$-linear isomorphism such that
$f(\Gamma/\Gamma\cap L)=\Gamma\cap L$ and let $L_f=\{f(v)+v\mid v\in
V/L\}$ be the graph of $f$, which is a linear subspace of $V$. Since
$f(\Gamma/\Gamma\cap L)=\Gamma\cap L$, the image $\overline{L}_f$ of
$L_f$ under the projection $V\lto V/\Gamma$ intersects each fiber of
$p:V/\Gamma\lto V/(L+\Gamma)$ in finitely many points. For the basis
$e_1,\cdots, e_d$ of $V=V/L\oplus L$, let $f(e_j)=\sum_{k=1}^g
f_{kj}e_{g+k}$. Since
\begin{align*}
\Omega(f(e_j)+e_j,f(e_i)+e_i)&=\Omega(f(e_j),e_i)+\Omega(e_j,f(e_i))\\
&=\Omega(\sum_kf_{kj}e_{g+k},e_i)-\Omega(\sum_kf_{ki}e_{g+k},e_j)\\
&=\sum_k\left\{f_{kj}\Z_{ki}-f_{ki}\Z_{kj}\right\}\\
&=(\Z^t f)_{ij}-(\Z^t f)_{ji},
\end{align*}
the graph $L_f$ of $f$ is an $\Omega$-Lagrangian subspace of $V$ if
and only if $\Z^tf:V/L\lto(V/L)^*$ is symmetric. Analogously, we
have the following relations:
\begin{equation}\label{lagr}
(\rZ)^tf=f^t(\rZ), \ \ \ (\iZ)^tf=f^t(\iZ).
\end{equation}

Associated to the linear map $f:V/L\lto L$, define an antisymmetric
bilinear form $E_f$ on $(\Gamma/\Gamma\cap L)\oplus (\Gamma\cap
L)^\perp$ as follows; for $u_1,u_2\in\Gamma/\Gamma\cap L$ and
$v_1^*,v_2^*\in (\Gamma\cap L)^\perp$,
\[E_f\left((u_1,v_1^*),(u_2,v_2^*)\right)=v_2^*\left(
f(u_1)\right)-v_1^*\left(f(u_2)\right).\] We extend $E_f$ to an
$\mathbb{R}$-bilinear anti-symmetric form on $V/L\oplus L^*$. Using
the identification $\rZ:V/L\cong L^*$, we consider the bilinear form
$E_f$ as the one on $V/L\oplus V/L$. Then the bilinear form can be
represented by
\[E_f=\begin{pmatrix}0&f^t(\rZ)\\-(\rZ)^tf&0\end{pmatrix}.\]
We now show that the graph $L_f$ of $f$ is an $\Omega$-Lagrangian
subspace of $V$ if and only if the bilinear form $E_f$ satisfies
$E_f(\widehat{J}_\Omega v,\widehat{J}_\Omega w)=E_f(v,w)$, for
$v,w\in V/L$. Suppose that $L_f$ is $\Omega$-Lagrangian, then the
relations (\ref{lagr}) hold. By the definition of
$\widehat{J}_\Omega$ given in (\ref{dcplex}), we have
\begin{equation}
\begin{pmatrix}-\mathsf{A}&-1\\1+\mathsf{A}^2&\mathsf{A}\end{pmatrix}^t
\begin{pmatrix}0&f^t(\rZ)\\-(\rZ)^tf&0\end{pmatrix}
\begin{pmatrix}-\mathsf{A}&-1\\1+\mathsf{A}^2&\mathsf{A}\end{pmatrix}:=
\begin{pmatrix}X&Y\\-Y^t&W\end{pmatrix},\nonumber
\end{equation}
where
\begin{align}
X&=-\A^tf^t(\rZ)(1+\A^2)+(1+\A^{2t})(\rZ)^tf\A \label{a}\\
Y&=-\A^tf^t(\rZ)\A+(1+\A^{2t})(\rZ)^tf\label{b}\\
W&=-f^t(\rZ)\A+\A^t(\rZ)^tf.\label{c}
\end{align}
since $\A=(\rZ)^{-1}\iZ$ and by (\ref{lagr}),
\begin{align}
W&=-f^t(\rZ)\A+\A^t(\rZ)^tf=-f^t\iZ+(\iZ)^tf=0\label{cor}
\end{align}
By (\ref{lagr}) and (\ref{cor}), we have
\begin{align*}
X&=-\A^tf^t(\rZ)(1+\A^2)+(1+\A^{2t})(\rZ)^tf\A\\
&=-\A^t(\rZ)^tf(1+\A^2)+(1+\A^{2t})f^t(\rZ)\A\\
&=-\A^t(\rZ)^tf\A^2+\A^{2t}f^t(\rZ)\A\\
&=-\A^t\left((\rZ)^tf\A+\A^{t}f^t(\rZ)\right)\A=0
\end{align*}
Using  (\ref{cor}) again, we have
\begin{align*}
Y&=-\A^tf^t(\rZ)\A+(1+\A^{2t})(\rZ)^tf\\
&=-\A^tf^t(\rZ)\A+\A^{2t}(\rZ)^tf+(\rZ)^tf\\
&=-\A^tf^t(\rZ)\A+\A^tf^t(\rZ)\A+(\rZ)^tf\\
&=(\rZ)^tf
\end{align*}
Thus, by (\ref{lagr}),
\begin{align}\label{d}
\begin{pmatrix}X&Y\\-Y^t&W\end{pmatrix}=\begin{pmatrix}0&f^t(\rZ)\\-(\rZ)^tf&0\end{pmatrix}
\end{align}
as desired. Conversely, suppose $E_f(\widehat{J}_\Omega
v,\widehat{J}_\Omega w)=E_f(v,w)$, then (\ref{d}) is true and we
show that the relations (\ref{lagr}) is also true. Since $W=0$ and
by (\ref{cor}), we have $f^t\iZ=\iZ^tf$. Also, $f^t\iZ=\iZ^tf$
implies that  $Y=(\rZ)^tf$. Thus, by the relation (\ref{d}), we
should have $f^t(\rZ)=(\rZ)^t f$.  Now, the graph $L_f$ is an
$\Omega$-Lagrangian subspace of $V$.

Finally, we shall show how to find a $\Omega$-Lagrangian submanifold
of $\mathbb{T}^d$ from a holomorphic line bundle over $(\T,\Omega)$.
This will allow us to compare the $\Omega$-Lagrangian property given
in (\ref{lagr}) with the Riemann conditions ({\it cf.} \cite{GH}).
Also, as we will see in the next section, the argument given here is
easily applied to the case of the noncommutative tori.

A holomorphic line bundle on $\T$ is specified by its first Chern
class which can be represented by an anti-symmetric bilinear form on
$\Lambda$. Let $E_f$ be any integral anti-symmetric bilinear form on
$W=V/L\oplus L^*$. Without loss of generality, we may assume that
$E_f$ is given by the matrix
$\begin{pmatrix}0&f^t\\-f&0\end{pmatrix}$ for the basis given above.
Note that we also regard $f$ as a linear map from $V/L$ to $L$.
Associated to such a matrix $E_f$, one can find a complex $g\times
g$ matrix $Z$ such that the $g\times d$ matrix
$\begin{pmatrix}Z&f^t\end{pmatrix}$ is a period matrix over the
lattice $\Lambda$. Now, the 2-form $E_f$ on $\T$ is of type $(1,1)$
if and only if
\[\begin{pmatrix}Z&f^t\end{pmatrix}\begin{pmatrix}0&-f^{-1}\\f^{-t}&0\end{pmatrix}
\begin{pmatrix}Z^t\\f\end{pmatrix}=0,\] which implies that $Z=Z^t$.
From the period matrix, we can reconstruct the complex structure
$\widehat{J}_\Omega$ in the same basis. Note that the complex
structure given in (\ref{dcplex}) is defined on $V/L\oplus V/L$
using the identification $\rZ:V/L\cong L^*$.  Thus we must consider
the period matrix using the basis for $V/L\oplus V/L$ and this is
done using $\rZ$. Now, in order to find the complex structure
$\widehat{J}_\Omega$, we need to solve the matrix system
\[\begin{pmatrix}Z&f^t\end{pmatrix}\widehat{J}_\Omega
=i\begin{pmatrix}Z&f^t\end{pmatrix}\]
or equivalently
\[\begin{pmatrix}\text{Re }Z&f^t\rZ\\\text{Im
}Z&0\end{pmatrix}\widehat{J}_\Omega=\begin{pmatrix}-\text{Im
}Z&0\\\text{Re }Z&f^t\rZ\end{pmatrix}.\] For simplicity, we let
$\mathsf{f}=f^t\cdot\rZ$. If we consider the matrix $f$ is regarded
as the linear map $f:V/L\lto L$, the $\mathsf{f}$ is regarded as the
linear map from $V/L$ to $\left(V/L\right)^*$, using the
identification $\rZ:V/L\cong L^*$. Now, we have
\begin{align}
\widehat{J}_\Omega&=\begin{pmatrix}\text{Re }Z&f^t\rZ\\\text{Im
}Z&0\end{pmatrix}^{-1}\begin{pmatrix}-\text{Im }Z&0\\\text{Re
}Z&f^t\cdot\rZ\end{pmatrix}
\nonumber\\
&=\begin{pmatrix}0&(\text{Im
}Z)^{-1}\\\mathsf{f}^{-1}&-\mathsf{f}^{-1}\text{Re }Z(\text{Im
}Z)^{-1}\end{pmatrix}\begin{pmatrix}-\text{Im }Z&0\\\text{Re
}Z&f^t\cdot\rZ\end{pmatrix}\nonumber\\
&=\begin{pmatrix}(\text{Im }Z)^{-1}\text{Re }Z&(\text{Im
}Z)^{-1}\mathsf{f}\\
-\mathsf{f}^{-1}\left[\text{Im }Z+\text{Re }Z(\text{Im
}Z)^{-1}\text{Re }Z\right]&-\mathsf{f}^{-1}\text{Re }Z(\text{Im
}Z)^{-1}\mathsf{f}
\end{pmatrix}\label{7}
\end{align}
The formula (\ref{7}) is also given in \cite{Man}. By comparing
(\ref{7}) and (\ref{dcplex}), we get the following consistency
relations:
\begin{align}
\mathsf{A}&=-(\text{Im }Z)^{-1}\text{Re }Z=-\mathsf{f}^{-1}\text{Re
}Z(\text{Im
}Z)^{-1}\mathsf{f}\label{1}\\
1&=-(\text{Im }Z)^{-1} \mathsf{f}\label{2}\\
1+\mathsf{A}^2&=-\mathsf{f}^{-1}\left[\text{Im }Z+\text{Re
}Z(\text{Im }Z)^{-1}\text{Re }Z\right]\label{3}
\end{align}
It is easy to check that the conditions (\ref{2}) and (\ref{3}) are
equivalent. By (\ref{2}), we  have
\begin{equation}\label{syf}
\mathsf{f}=-\text{Im }Z
\end{equation}
and also by (\ref{1}) and (\ref{syf})
\begin{equation*}
\mathsf{f}\mathsf{A}=-\mathsf{f}(\text{Im }Z)^{-1}\text{Re
}Z=-\text{Re }Z(\text{Im }Z)^{-1}\mathsf{f}=\text{Re }Z.
\end{equation*}
Thus we see that if $Z=Z^t$, then $\mathsf{f}$ and $\mathsf{fA}$ are
symmetric. Similarly, if $Z=\overline{Z}^t$, then $\mathsf{f}$ is
skew-symmetric and $\mathsf{fA}$ is symmetric.

Conversely, suppose that $\mathsf{f}$ and $\mathsf{fA}$ are
symmetric. Then by (\ref{syf}), $\iiZ$ is symmetric. From (\ref{1})
and (\ref{2}), we have
$\mathsf{fA}=-\mathsf{f}(\iiZ)^{-1}\rrZ=\rrZ$. Thus if $\mathsf{fA}$
is symmetric, then $\rrZ$ is symmetric, which implies that $Z=Z^t$.
Also, we see that  $\iiZ>0$ if and only if $\mathsf{f}<0$, by
(\ref{2}). Thus, we see that the Riemann conditions on the complex
side is the mirror dual to the $\Omega$-Lagrangian property given in
(\ref{lagr}). On the other hand, if $\mathsf{f}$ is skew-symmetric
and $\mathsf{fA}$ is symmetric, then we have $Z=\overline{Z}^t$.
Thus, in this case, we see that a non-holomorphic bundle on the
mirror dual torus corresponds to a submanifold of the original torus
which is defined to be the graph of a skew-symmetric linear map
after the identification with $\rZ:V/L\cong L^*$.

In particular, an interesting fact is that the correspondence
between complexified symplectic form on $\mathbb{T}^d$ and the
complex structure on the mirror dual torus is easily seen by the
relations
\begin{equation}\label{mirror}
\mathsf{f}=f^t\rZ=-\iiZ, \ \ \ \ \mathsf{fA}:=f^t\iZ=\rrZ
\end{equation}
and hence
\begin{align*}
f^t\mathcal{Z}&=f^t\cdot\rZ+if^t\cdot\iZ\\
&=-\iiZ+i\rrZ\\
&=i(\rrZ+i\iiZ)=iZ
\end{align*}

As a conclusion, the mirror dual complex torus
$(\T,\widehat{J}_\Omega)$, equipped with the integral 2-form $E_f$,
is an abelian variety if and only if the real matrices $\mathsf{f}$,
$\mathsf{f}\mathsf{A}$ are symmetric and $\mathsf{f}<0$. This might
be understood as a symplectic version of the Riemann condition.
Also, for a holomorphic line bundle $\widehat L$ on $\T$ such that
$c_1(\widehat{L})\in H^{1,1}(\T,\mathbb{R})\cap H^2(\T,\mathbb{Z})$,
we may write $c_1(\widehat{L})$ as an integral bilinear form
$E_f=\begin{pmatrix}0&f^t\\-f&0\end{pmatrix}$ on $W=V/L\oplus L^*$.
Then the graph of the integral linear map $f:V/L\to L$ is an
$\Omega$-Lagrangian subspace of $V$. This analysis will be
generalized to the case of noncommutative  complex tori in the next
section.


\section{Mirror duality on generalized complex tori}

The aim of this section is to rephrase the mirror duality given in
Section 2 in terms of generalized complex structures which were
introduced by Hitchin (\cite{Hit} and see also \cite{Gual}).
Modifying the notion ``T-duality in all direction'' defined by
Kapustin in \cite{Kap}, we define T-duality in half direction and we
will show that the duality is well matched with the mirror symmetry
given in Section 2.

Let us first recall the definition of a generalized complex
structure on a real vector space. Let $V$ be a $d$-dimensional
vector space over $\mathbb{R}$. Then the space $V\oplus V^*$ is
naturally equipped with a pseudo-Euclidean metric defined by
\begin{equation}\label{gmetric}
\langle
X+v^*,Y+w^*\rangle=\frac{1}{2}\left(v^*(Y)+w^*(X)\right)=\frac{1}{2}\begin{pmatrix}
X&v^*\end{pmatrix}\begin{pmatrix}0&1\\1&0\end{pmatrix}\begin{pmatrix}Y\\w^*\end{pmatrix},
\end{equation}
for $X,Y\in V$ and $v^*,w^*\in V^*$. For simplicity we write
$q=\begin{pmatrix}0&1\\1&0\end{pmatrix}$ as the pseudo-Euclidean
metric given in (\ref{gmetric}). A {\it generalized complex
structure} on $V$ is an endomorphism $\mathcal{J}$ of $V\oplus V^*$
satisfying $\mathcal{J}^2=-1$ and $\mathcal{J}\mathcal{J}^t=-1$,
i.e., $\mathcal{J}$ is orthogonal with respect to the
pseudo-Euclidean metric. This orthogonality can be seen as the
following commuting diagram:
\[\begin{CD}
V^*\oplus V@>\mathcal{J}^t>> V^*\oplus V\\
@VqVV  @V
qVV\\
V\oplus V^*@>-\mathcal{J}>>V\oplus V^*
\end{CD}\]

A generalized complex torus
$(\mathbb{T}^d,\mathcal{J}_1,\mathcal{J}_2)$ is a real torus
$\mathbb{T}^d=V/\Gamma$ equipped with a pair
$(\mathcal{J}_1,\mathcal{J}_2)$ of generalized complex structures on
$\mathbb{T}^d$ such that
$\mathcal{J}_1\mathcal{J}_2=\mathcal{J}_2\mathcal{J}_1$ and
$G=-\mathcal{J}_1\mathcal{J}_2$ is a positive definite metric on
$V\oplus V^*$. In particular, such a pair of generalized complex
structures is called a generalized K\"ahler structure on
$\mathbb{T}^d$. A typical example is given as follows: Let $J\in
\text{End}(V)$ be a complex structure on $\mathbb{T}^d=V/\Gamma$
endowed with a constant K\"ahler form $\omega$ with a B-field $\xi$,
and with a flat Riemannian metric $g$. Then we have two generalized
complex structures:
\[\mathcal{J}_J=\begin{pmatrix}J&0\\0&-J^t\end{pmatrix}, \ \ \
\mathcal{J}_\omega=\begin{pmatrix}0&-\omega^{-1}\\\omega&0\end{pmatrix}\]
and we may transform them by a B-field $\xi$:
\begin{align*}
\mathcal{J}_J^\xi&=\begin{pmatrix}1&0\\\xi&1\end{pmatrix}
\begin{pmatrix}J&0\\0&-J^t\end{pmatrix}\begin{pmatrix}1&0\\-\xi&1\end{pmatrix}
=\begin{pmatrix}J&0\\\xi J+J^t\xi&-J^t\end{pmatrix}\\
\mathcal{J}_\omega^\xi&=\begin{pmatrix}1&0\\\xi&1\end{pmatrix}
\begin{pmatrix}0&-\omega^{-1}\\\omega&0\end{pmatrix}\begin{pmatrix}1&0\\-\xi&1\end{pmatrix}
=\begin{pmatrix}\omega^{-1}\xi&-\omega^{-1}\\\omega+\xi\omega^{-1}\xi&-\xi\omega^{-1}\end{pmatrix}.
\end{align*}
It is easy to check that
$(\mathbb{T}^d,\mathcal{J}_J^\xi,\mathcal{J}_\omega^\xi)$ is a
generalized complex torus. Now, such two generalized complex tori
$(\mathbb{T}^d_1,\mathcal{J}_{J_1}^{\xi_1},\mathcal{J}_{\omega_1}^{\xi_1})$
and
$(\mathbb{T}^d_2,\mathcal{J}_{J_2}^{\xi_2},\mathcal{J}_{\omega_2}^{\xi_2})$
are mirror of each other if there is a lattice isomorphism
$\phi:\Gamma_1\oplus\Gamma_1^*\lto \Gamma_2\oplus\Gamma_2^*$ such
that $\phi^tq_2\phi=q_1$ and
$\phi^{-1}\mathcal{J}_{J_1}^{\xi_1}\phi=\mathcal{J}_{\omega_2}^{\xi_2}$,
$\phi^{-1}\mathcal{J}_{\omega_1}^{\xi_1}\phi=\mathcal{J}_{J_2}^{\xi_2}$,
where $q_i=\begin{pmatrix}0&1\\1&0\end{pmatrix}$, $i=1,2$, are the
pseudo-Euclidean metric.

We now rephrase the mirror duality given in Section 2 in terms of
generalized K\"ahler structure by constructing an explicit mirror
map $\phi$ and the map will be referred as a T-duality in half
direction(compare with \cite{Kap}). Let
$\mathbb{T}^d=V/\Gamma=\mathbb{C}^g/(\mathbb{Z}\oplus i\mathbb{Z})$
be a complex torus equipped a complexfied symplectic form
$\Omega=\omega+i\xi$. Then the mirror of
$(\mathbb{T}^d=V/\Gamma,\Omega)$ is given by $W/\Lambda$, where
$W=(V/L)\oplus L^*$ and $\Lambda=(\Gamma/\Gamma\cap
L)\oplus(\Gamma\cap L)^\perp$ by decomposing
$\Gamma=(\Gamma/\Gamma\cap L)\oplus (\Gamma\cap L)$. We define
\begin{equation}\label{mirrormap}
\phi=\begin{pmatrix}-1&0&0&0\\0&0&0&1\\0&0&-1&0\\0&1&0&0\end{pmatrix}:\Gamma\oplus\Gamma^*\lto\Lambda\oplus\Lambda^*.
\end{equation}
We shall verify the map $\phi$ gives the mirror correspondence
discussed in Section 2.  Note that  we have a generalized K\"ahler
structure on $\mathbb{T}^d=V/\Gamma$ is given by
\begin{align}
\mathcal{J}_J^\xi&=\begin{pmatrix}0&-1&0&0\\1&0&0&0\\0&0&0&-1\\0&0&1&0\end{pmatrix}\label{J1}\\
\mathcal{J}_\omega^\xi&=\begin{pmatrix}\omega^{-1}\xi&-\omega^{-1}\\
\omega+\xi\omega^{-1}\xi&-\xi\omega^{-1}\end{pmatrix}\label{J2}
\end{align}
The equation (\ref{J1}) follows from the fact that $\xi$ is a type
of $(1,1)$ with respect to the canonical complex structure
$J=\begin{pmatrix}0&-1\\1&0\end{pmatrix}$. Using the notation given
in Section 2, the entries in  (\ref{J2}) are given as follows:
\begin{align*}
\omega^{-1}\xi&=\begin{pmatrix}(\rZ)^{-1}\iZ&0\\0&(\rZ)^{-t}\iZ^t\end{pmatrix}\\
-\omega^{-1}&=\begin{pmatrix}0&-(\rZ)^{-1}\\(\rZ)^{-t}&0\end{pmatrix}\\
\omega+\xi\omega^{-1}\xi&=\begin{pmatrix}0&-\rZ^t-\iZ^t(\rZ)^{-t}\iZ^t\\\rZ+\iZ(\rZ)^{-1}\iZ&0\end{pmatrix}\\
-\xi\omega^{-1}&=\begin{pmatrix}-\iZ^t(\rZ)^{-t}&0\\0&-\iZ(\rZ)^{-1}\end{pmatrix}.
\end{align*}
Then it is easy to compute
\begin{align*}
\phi^{-1}\mathcal{J}_J^\xi\phi=\begin{pmatrix}0&0&0&-1\\0&0&1&0\\0&-1&0&0\\1&0&0&0\end{pmatrix}
\end{align*}
and
\begin{align*}
\phi^{-1}\mathcal{J}_\omega^\xi\phi=-\begin{pmatrix}\widehat{J}_\Omega&0\\0&-\widehat{J}_\Omega^t\end{pmatrix}
\end{align*}
where $\widehat{J}_\Omega$ is the complex structure on the mirror
dual torus given in Section 2.

Using the mirror map defined above, we may consider the case that
the complex 2-form $\Omega=\omega+i\xi$ is not of type (1,1). Since
we have $\xi$ as a B-field, we shall keep $\omega$ as of type (1,1).
With the same basis $e_1,\cdots,e_g,e_{g+1},\cdots,e_d$ for the
vector space $V$ given in Section 2, we define a complexified
symplectic form $\Omega$ as follows;
$\Omega(e_i,e_j)=\sqrt{-1}X_{ij}$, $\Omega(e_{g+i},e_j)=\Z_{ij}$,
$\Omega(e_{g+i},e_{g+j})=0$, where $X_{ij}$ is  real. Then we may
represent $\xi$ as a block matrix
$\xi=\begin{pmatrix}X&-\iZ^t\\\iZ&0\end{pmatrix}$. In other words,
we only have common $\omega$ and $\xi$-Lagrangian subspaces on the
base space of the Lagrangian torus fibrations considered in Section
2. One finds that $\xi$ is no more of type $(1,1)$.  To be more
precise, recall we have a canonical complex structure
$\begin{pmatrix}0&-1\\1&0\end{pmatrix}$ and consider the following
general form
\[\begin{pmatrix}0&1\\-1&0\end{pmatrix}\begin{pmatrix}X&-\iZ^t\\-\iZ&Y\end{pmatrix}
\begin{pmatrix}0&-1\\1&0\end{pmatrix}=\begin{pmatrix}Y&-\iZ\\\iZ^t&X\end{pmatrix}.\]
From the relation above, we see that the 2-form which is represented
by the matrix $\begin{pmatrix}X&-\iZ^t\\-\iZ&Y\end{pmatrix}$ is of
type $(1,1)$ if and only if $X=Y$ and the matrix $\iZ$ is symmetric.
Also, $\xi$ is of type $(2,0)$ or $(0,2)$ if and only if $X=-Y$ and
$\iZ$ is anti-symmetric. Then since
\[\xi=\begin{pmatrix}X&-\iZ^t\\\iZ&0\end{pmatrix}=\begin{pmatrix}\frac{X}{2}&
\frac{\iZ-\iZ^t}{2}\\\frac{\iZ-\iZ^t}{2}&-\frac{X}{2}\end{pmatrix}+
\begin{pmatrix}\frac{X}{2}&
-\frac{\iZ+\iZ^t}{2}\\\frac{\iZ+\iZ^t}{2}&\frac{X}{2}\end{pmatrix},\]
we see that the $\xi$ is a most general type of B-field. Now the
generalized K\"ahler structure is given by
\begin{align}
\mathcal{J}_J^\xi&=\begin{pmatrix}J&0\\\xi J+J^t\xi&-J^t\end{pmatrix}\label{Jj1}\\
\mathcal{J}_\omega^\xi&=\begin{pmatrix}\omega^{-1}\xi&-\omega^{-1}
\\\omega+\xi\omega^{-1}\xi&-\xi\omega^{-1}\end{pmatrix}\label{Jj2}
\end{align}
and
\begin{align*}
\omega^{-1}\xi&=\begin{pmatrix}(\rZ)^{-1}\iZ&0\\-(\rZ)^{-t}X&(\rZ)^{-t}\iZ^t\end{pmatrix}\\
-\omega^{-1}&=\begin{pmatrix}0&-(\rZ)^{-1}\\
(\rZ)^{-t}&0\end{pmatrix}\\
\omega+\xi\omega^{-1}\xi=&
\begin{pmatrix}(\iZ)^t(\rZ)^{-t}X+X(\rZ)^{-1}
\iZ&-\rZ^t-\iZ^t(\rZ)^{-t}\iZ^t\\\rZ+\iZ(\rZ)^{-1}\iZ&0\end{pmatrix}\\
-\xi\omega^{-1}&=\begin{pmatrix}-\iZ^t(\rZ)^{-t}&-X(\rZ)^{-1}\\0&-\iZ(\rZ)^{-1}\end{pmatrix}.
\end{align*}

For the complex structure (\ref{Jj1}), since
\begin{align*}
\xi
J+J^t\xi&=\begin{pmatrix}X&-\iZ^t\\\iZ&0\end{pmatrix}\begin{pmatrix}0&-1\\
1&0\end{pmatrix}+\begin{pmatrix}0&1\\-1&0\end{pmatrix}
\begin{pmatrix}X&-\iZ^t\\\iZ&0\end{pmatrix}\\
&=\begin{pmatrix}\iZ-\iZ^t&-X\\-X&\iZ^t-\iZ\end{pmatrix},
\end{align*}
we have
\[\mathcal{J}_J^\xi=\begin{pmatrix}0&-1&0&0\\1&0&0&0\\\iZ-\iZ^t&-X&0&-1\\-X&\iZ^t-\iZ&1&0\end{pmatrix}\]
By applying the mirror map, we get
\[\phi^{-1}\mathcal{J}_J^\xi\phi=\begin{pmatrix}0&0&0&1\\X&0&-1&\iZ^t-\iZ\\\iZ-\iZ^t&1&0&X\\-1&0&0&0\end{pmatrix}\]
Similarly, we have
\begin{align*}
\phi^{-1}\mathcal{J}_\omega^\xi\phi=\begin{pmatrix} A&B\\C&D
\end{pmatrix}
\end{align*}
where
\begin{align*}
A&=\begin{pmatrix}
(\rZ)^{-1}\iZ&(\rZ)^{-1}\\-\rZ-\iZ(\rZ)^{-1}\iZ&-\iZ(\rZ)^{-1}\end{pmatrix}\\
B&=\begin{pmatrix}0&0\\0&0\end{pmatrix}\\
C&=\begin{pmatrix}(\iZ)^t(\rZ)^{-t}X+X(\rZ)^{-1}\iZ&X(\rZ)^{-1}\\(\rZ)^{t}X&0\end{pmatrix}\\
D&=\begin{pmatrix}-\iZ^t(\rZ)^{-t}&\rZ^t+\iZ^t(\rZ)^{-t}\iZ^t\\-(\rZ)^{-t}&(\rZ)^{-t}\iZ^t\end{pmatrix}.
\end{align*}
Finally, we note that the B-field $\xi$ is mapped to
$\phi\xi\phi^{-1}$ under the mirror map $\phi$ and we get the above
correspondence.

Note that the generalized K\"ahler structure $\mathcal{J}_J^\xi$
given in (\ref{Jj1}) is block-lower triangular. Thus if we take
``T-duality in all directions'' as defined in \cite{Kap}, the
corresponding generalized K\"ahler structure on the dual torus
becomes a block-upper triangular. However, this is impossible unless
the B-field $\xi$ is of type (1,1) related to a $N=2$ super
conformal field theory. Hence if the (0,2)-part of $\xi$ is not 0,
then the T-duality has a problem. Analogously, since T-duality
between a complex torus and its dual torus makes the categories of
B-branes on both tori  equivalent, the category of B-branes should
be twisted by the (0,2)-part of $\xi$. From this, Kapustin proposed
that such a twistedness is characterized by the fact that the matrix
$\mathcal{J}_J^\xi$ is block-lower-triangular, see \cite{Kap} for
details.

We have chosen a specific type of B-field which has a nonzero
(0,2)-part and  it  defines a twistedness on B-brane category on the
complex torus as discussed above. On the other hand,  from our
choice of the B-field $\xi$, we find that the A-brane category on
$\mathbb{T}^d$ should be deformed by the (0,2)-part of $\xi$ since
the generalized K\"ahler structure $\mathcal{J}_\omega^\xi$ given in
(\ref{Jj2}) is block-lower-triangular after taking a T-duality in
half directions as shown in the above computations. As a conclusion,
we state a symplectic version of Kapustin's proposal: a
noncommutative deformations are characterized by the fact that the
mirror dual of symplectic type of generalized K\"ahler structure
such as the one given in (\ref{Jj2}), is block-lower-triangular.
From the categorical point of view, the A-brane category is twisted
by the (0,2)-part of the given B-field.


\section{Mirror duality on  noncommutative complex tori}

In this section, we generalize the mirror duality on abelian
varieties to the case of noncommutative complex tori. Let us first
recall some basic facts for a noncommutative complex torus and
holomorphic structures on it, see \cite{Sch01} for details. A
noncommutative torus $\T_\theta$  is generated by $d$-unitaries
$U_1,\cdots,U_d$
\begin{equation}\label{co}
U_iU_j=\exp(2\pi i\theta_{ij})U_jU_i,
\end{equation}
where $\theta=(\theta_{ij})$ is an irrational $d\times d$
skew-symmetric matrix. The relation (\ref{co}) defines the
presentation of the involutive algebra
\[A_\theta^d=\left\{
\sum_{(n_1,\cdots,n_d)\in\mathbb{Z}^d}a_{n_1,\cdots,n_d}U_1^{n_1}\cdots
U_d^{n_d}\mid a_{n_1,\cdots,n_d}\in
\mathcal{S}(\mathbb{Z}^d)\right\}\] where the coefficient function
$(n_1,\cdots,n_d)\mapsto a_{n_1,\cdots,n_d}$ rapidly decays at
infinity. By definition, the algebra $A_\theta^d$ is the algebra of
smooth functions on $\T_\theta$. The ordinary torus $\T$ acts on the
algebra $A_\theta^d$ ({\it cf.} \cite{Ri1}) and the infinitesimal
form of the action of $\T$ on $A_\theta^d$ defines a Lie algebra
homomorphism
\begin{equation}\label{der}
\delta:W\lto \text{ Der}(A^d_\theta),
\end{equation}
where $\text{ Der}(A^d_\theta)$ denotes the Lie algebra of
derivations of $A_\theta^d$. Generators $\delta_1,\cdots, \delta_d$
of $\text{Der}(A^d_\theta)$ act as follows:
\[\delta_j(U_j)=2\pi
iU_j \text{ \ \ and \ \ }\delta_i(U_j)=0 \text{ for } i\ne j.\]

A noncommutative torus $\T_\theta$ is said to be a {\it
noncommutative complex torus} if the Lie algebra $W\cong
\mathbb{R}^d$ is equipped with a complex structure.  Associated to a
given complex structure on $W$, the complexification
$W\otimes_\mathbb{R}\mathbb{C}$ can be decomposed by two complex
conjugate subspaces $W^{0,1}$ and $W^{1,0}$, which are of complex
dimension $g$. Let
\[\Omega^{0,p}=\wedge^p(W^{0,1})^* \text{ \ \ and \ \ }
\Omega^{0,\bullet}=\bigoplus_{p=0}^g\Omega^{0,p}.\] A holomorphic
structure on a vector bundle $\mathcal{E}$ over $\T_\theta$, which
corresponds to a finitely generated projective (left)
$A_\theta$-module, is given by a linear map
\[\overline{\nabla}:\mathcal{E}\otimes\Omega^{0,\bullet}\lto
\mathcal{E}\otimes \Omega^{0,\bullet+1}\] satisfying
\begin{equation}\label{nomeaning}
\overline{\nabla}_\alpha(u\cdot e)=u\cdot\overline{\nabla}_\alpha
e+\bar\delta_\alpha(u)\cdot e, \ \ \ u\in A_\theta, \
e\in\mathcal{E}\end{equation} and
\begin{equation}\label{dg}
[\overline{\nabla}_\alpha,\overline{\nabla}_\beta]=0,
\end{equation}
where $\bar\delta_1,\cdots,\bar\delta_g$ are generators for the Lie
algebra $\text{ Der}(A^d_\theta)$ associated to a basis for
$W^{0,1}$. From the condition (\ref{dg}), we get a complex
\[\begin{CD}
0@>>>\mathcal{E}@>\overline{\nabla}>>\mathcal{E}\otimes(W^{0,1})^*
@>\overline{\nabla}>>\mathcal{E}\otimes\wedge^2(W^{0,1})^*@>>>\cdots,
\end{CD}\]
and the corresponding cohomology will be denoted by
$H^*(\mathcal{E},\overline{\nabla})$.  Note that, since
$\text{dim}_\mathbb{C}W^{0,1}=g$,
$H^k(\mathcal{E},\overline{\nabla})=0$ if $k>g$. In particular,
$H^0(\mathcal{E},\overline{\nabla})$ consists of
$\phi\in\mathcal{E}$ such that $\overline{\nabla}\phi=0$. The
elements of $H^0(\mathcal{E},\overline{\nabla})$ are called {\it
holomorphic vectors} or {\it theta vectors}.

The vector bundles over $\T$ are classified by the $K$-theory of
$\T$ and there is a ring homomorphism $\text{ch}:K^*(\T)\lto
H^*(\T,\mathbb{Q})$. Similarly, finitely generated projective
$A_\theta^d$-modules are classified by $K_0(A_\theta^d)$ and the
Chern character takes values in $H^*(\T,\mathbb{R})$. The targets of
the both Chern characters are related by the deformation parameter
$\theta\in\wedge^2W$. The relation is summarized by the following
diagram:
\[\begin{CD}K^0(\T)@>\text{ch}>>H^{\text{even}}(\T,\mathbb{Q})\\
& & @V e^{i(\theta)}VV\\
K_0(A_\theta^d)@>\text{Ch}>>H^{\text{even}}(\T,\mathbb{R})
\end{CD}\]
where $i(\theta)$ denotes the contraction with 2-vector $\theta$.
Thus, for a given vector bundle $E$ over $\T$, one can construct an
$A_\theta^d$-module $\mathcal{E}$ such that
\begin{equation}\label{chern}
\text{Ch}(\mathcal{E})=e^{i(\theta)}\text{ch}(E).
\end{equation}
Note that  the cohomology group $H^\bullet(\T,\mathbb{R})$ can be
identified with the exterior algebra $\wedge^\bullet W^*$, where
$W^*=\text{Hom}_\mathbb{R}(W,\mathbb{R})$ is the dual vector space
of $W$. In below, we shall study the mirror dual property of the
cohomological deformation described above.

Let $f:V/L\lto L$ be an integral linear map and let
$L_f=\{f(v)+v\mid v\in V/L\}$ be the graph of $f$. Since $f$ is
integral, $L_f\cap \Gamma\cong\mathbb{Z}^g$ and
$\overline{L}_f=L_f/(L_f\cap\Gamma)$ intersects each fiber of
$p:V/\Gamma\lto V/(L+\Gamma)$ in one point. As we have discussed in
Section 2, the linear subspace $L_f$ defines an integral
antisymmetric bilinear form
$E_f=\begin{pmatrix}0&f^t\\-f&0\end{pmatrix}$ on $W=V/L\oplus L^*$.
Then the linear subspace $L_f$ of $V$ is an $\Omega$-Lagrangian if
and only if $E_f$ can be seen as an element in
$H^{1,1}(\T,\mathbb{C})\cap H^2(\T,\mathbb{Z})$. Then there is a
holomorphic line bundle $\widehat{L}_f$ on $\T$ such that
$c_1(\widehat{L}_f)=E_f$. Let us denote by $\#(\overline{L}_f\cap
\overline{L})$ the intersection number of Lagrangians
$\overline{L}_f$ and $\overline{L}$ in $V/\Gamma$. Since
$\#(\overline{L}_f\cap \overline{L})=1$, it is easy to see that
\[\text{Pf }E_f=\#(\overline{L}_f\cap \overline{L'}),\]
where $\text{Pf }E_f$ is the Pfaffian of the anti-symmetric form
$E_f$ and $\overline{L'}$ is the image of $V/L$ under the covering
map $V\lto V/\Gamma$. Note that the moduli space of the flat
Lagrangian submanifolds of $\mathbb{T}^d$ parallel to
$L_f/L_f\cap\Gamma$ is identified with the $d$-dimensional torus.

Based on the construction given in \cite{Ri1}, we shall deform the
line bundle $\widehat{L}_m$ on $\T$ to a holomorphic bundle over the
noncommutative torus $\T_\theta$. A finitely generated projective
$A^d_\theta$-module, which is in fact  a bundle over $\T_\theta$, is
given by a Schwarz space $\mathcal{S}(\mathbb{R}^g\times G)$, where
$G$ is a finite abelian group. Let $G=\prod_{i=1}^g\mathbb{Z}_{m_i}$
and let $\mathcal{E}=\mathcal{S}(\mathbb{R}^g\times G)$, where
$m_1m_2\cdots m_g=\text{Pf }E_f$ corresponds to the degree of the
line bundle $\widehat{L}_f$. Using the representation of the
Heisenberg commutation relations for the finite group $G$, one can
find unitary operators $W_i$ acting on $\mathcal{S}(G)=
\mathbb{C}^{m_1}\otimes \cdots \otimes \mathbb{C}^{m_g}$ such that
\begin{equation}\label{rational}
W_iW_j=\exp[2\pi i(E_f^{-1})_{ij}]W_jW_i.
\end{equation}
The operators $W_j$ can also be obtained using the twist eating
solution studied in \cite{geba}, and such operators specify the line
bundle $\widehat{L}_f$. In order to define an $A_\theta^d$-module
action on $\mathcal{E}=\mathcal{S}(\mathbb{R}^g\times G)$, one needs
to consider an embedding of the lattice $\Lambda$ into
$\mathbb{R}^g\times(\mathbb{R}^g)^*$ in the sense of \cite{Ri1}.
Such an embedding map can be given by a real invertible $d\times d$
matrix $T$ satisfying the relation
\begin{equation}\label{embedding}
T\begin{pmatrix}0&1\\-1&0\end{pmatrix}T^t=\gamma, \end{equation}
where $\gamma$ is an irrational skew-symmetric matrix such that
$E_f^{-1}-\gamma=\theta$. Let us denote by $T_{[1,j]}$, $T_{[2,j]}$
the first and the second $g$-rows in the $j$-th column of the matrix
$T$, respectively. Then associated to the embedding $T$, we define
operators $V_j$ on $\mathcal{S}(\mathbb{R}^g)$ by
\begin{equation}\label{real}
(V_jh)(\mathbf{s})=\exp\left(2\pi i\mathbf{s}^t\cdot
T_{[2,j]}\right) h(\mathbf{s}+T_{[1,j]}^t)
\end{equation}
where $\mathbf{s}=\begin{pmatrix} s_1& \cdots &
s_g\end{pmatrix}^t\in \mathbb{R}^g$ and
$h\in\mathcal{S}(\mathbb{R}^g)$. Then the operators satisfy the
following commutation relation
\begin{equation}\label{real2}
V_iV_j=\exp(-2\pi i\gamma_{ij})V_jV_i.
\end{equation}
Combining (\ref{rational}) and (\ref{real2}), the unitary operators
$U_i=V_i\otimes W_i$ defines an $A_\theta^d$-module action on
$\mathcal{E}$. Thus we get a vector bundle $\mathcal{E}$ on the
noncommutative torus $\T_\theta$. Similarly, one can construct a
constant curvature connection $\nabla$ on $\mathcal{E}$ using the
inverse matrix $T^{-1}$. More explicitly, define
\begin{equation}\label{connection}
(\nabla_jh)(\mathbf{s})=2\pi i\mathbf{s}^t\cdot T^{-1}_{[1,j]}
h(\mathbf{s})-\frac{\partial h}{\partial \mathbf{s}}\cdot
T^{-1}_{[2,j]},
\end{equation}
where $\frac{\partial h}{\partial \mathbf{s}}=\begin{pmatrix}
\frac{\partial h}{\partial s_1}& \cdots &\frac{\partial h}{\partial
s_g}\end{pmatrix}$. Then we have
\begin{equation}\label{curvature}
[\nabla_i,\nabla_j]=2\pi i(\gamma^{-1})_{ij}.
\end{equation}
Note that the operators of the form $\nabla_i+R_i$,
$R_i\in\mathbb{R}$ also satisfy the commutation relation
(\ref{curvature}) and thus the moduli space of such connections is a
$d$-dimensional torus.

Let us consider the case when the curvature $\gamma^{-1}$  is given
by the following simple block matrix:
\begin{equation}\label{simple}
\gamma^{-1}=\begin{pmatrix}0&F^t_{\gamma^{-1}}\\-F_{\gamma^{-1}}&0\end{pmatrix}
\ \ \text{ and } \ \
\gamma=\begin{pmatrix}0&-F_{\gamma}\\F^t_{\gamma}&0\end{pmatrix}
\end{equation}
where $F_{\gamma}$ is a $g\times g$ real matrix such that at least
one of its entries is irrational and
$F_{\gamma^{-1}}=F_{\gamma}^{-1}$. Note that the matrix $F_\gamma$
depends on the choice of a basis for $W$. Associated to the
curvature $\gamma^{-1}$ as given above, we define an antisymmetric
bilinear form $E_{\gamma^{-1}}$ on $W=V/L\oplus L^*$ as follows; for
$v_1,v_2\in V/L$ and $l_1^*, l_2^*\in L^*$,
\begin{equation}\label{realf}
E_{\gamma^{-1}}\left((v_1,l_1^*),(v_2,l_2^*)\right)
=l_2^*(f_{\gamma^{-1}}(v_1))-l_1^*(f_{\gamma^{-1}}(v_2))
\end{equation}
where $f_{\gamma^{-1}}:V/L\lto L$ is a linear map whose matrix is
given by $F_{\gamma^{-1}}$ with respect to the given basis for $W$.
Thus the curvature $\gamma^{-1}$ is the corresponding matrix for the
antisymmetric form $E_{\gamma^{-1}}$ on $W$. Also, we may regard the
antisymmetric form $E_{\gamma^{-1}}$ on $W$ as an element of the
cohomology group $H^2(\T,\mathbb{R})$ in the following way. Since
$\text{Ch}(\mathcal{E})=e^{i(\theta)}\text{ch}(\widehat{L}_f)\in
H^{2\bullet}(\T,\mathbb{R})$, we may write
\[\text{Ch}(\mathcal{E})=\text{Ch}_0(\mathcal{E})+\text{Ch}_1(\mathcal{E})+\cdots\in
H^0(\T,\mathbb{R})\oplus H^2(\T,\mathbb{R})\oplus\cdots\] The
curvature $E_{\gamma^{-1}}$ of $\nabla$ is given by $2\pi i
\text{Ch}_1(\mathcal{E})/\text{Ch}_0(\mathcal{E})$ and the 0-th
cohomology is computed to be
\[\text{Ch}_0(\mathcal{E})=\dim (\mathcal{E})=\det
T=\text{Pf}(\gamma).\] The first cohomology class is now given by
$\text{Ch}_1(\mathcal{E})=\dim (\mathcal{E})E_{\gamma^{-1}}\in
H^2(\T,\mathbb{R})=\wedge^2W^*$.  Thus, the curvature
$E_{\gamma^{-1}}$, which is normalized by the dimension of
$\mathcal{E}$, can be understood as an anti-symmetric form on
$W=V/L\oplus L^*$ representing $\text{Ch}_1(\mathcal{E})$.

Recall that the vector space $W$ is equipped with the complex
structure $\widehat{J}_\Omega$, which is defined in Section 2. The
linear map $\widehat{J}_\Omega$ is defined on $V/L\oplus V/L$ and by
identifying $V/L$ with $L^*$ via $\rZ$, it defines a complex
structure on $W$. Now by definition,
$(\T_\theta,\widehat{J}_\Omega)$ is a noncommutative complex torus.
We shall define a holomorphic structure on the vector bundle
$\mathcal{E}$ on $\T_\theta$ which is compatible with the complex
structure $\widehat{J}_\Omega$. Let
$W\otimes_\mathbb{R}\mathbb{C}=W^{1,0}\oplus W^{0,1}$, where
$W^{1,0}$ and $W^{0,1}$ are $i$ and $-i$ eigenspaces of
$\widehat{J}_\Omega$, respectively. Associated to the integral
anti-symmetric bilinear form $E_f$, which is the First Chern class
of the line bundle $\widehat{L}_f$, one has a period matrix of
$\Lambda\subset W$, as discussed in Section 2. Along the deformation
of the Chern character of $\widehat{L}_f$, we also deform the period
matrix. More explicitly, since $\text{Ch}_1(\mathcal{E})\in\wedge^2
W^*$, for the given basis for $W$, we may write
\[\text{Ch}_1(\mathcal{E})=\frac{1}{2}\sum_{i<j}\text{Pf}(\gamma)\left(F^t_{\gamma^{-1}}\right)_{ij}dx_i\wedge
dx_j\] where $x_1,\cdots, x_d$ are the dual coordinates on $W$. Then
the period matrix of $\Lambda\subset W$ is deformed to
$\mho_\gamma=\begin{pmatrix}F^t_\gamma&\sZ\end{pmatrix}$, where
$\sZ$ is a complex $g\times g$ matrix. By this period-like matrix,
we make the change of the basis for $W$ into the one for $W^{0,1}$.
Thus we let
\[\bar\delta=\mho_\gamma\delta \text{ \ \ or \ \
}\bar\delta_a=\sum_{j=1}^d\left(\mho_\gamma\right)_{aj}\delta_j.\]
According to the basis change, we define a holomorphic connection
$\overline{\nabla}$ on $\mathcal{E}$ by
$\overline{\nabla}=\mho_\gamma\nabla$. Then
\begin{align*}
[\overline{\nabla},\overline{\nabla}]&=[\mho_\gamma\nabla,\mho_\gamma\nabla]\\
&=\mho_\gamma {\gamma^{-1}}\mho_\gamma^t\\
&=\begin{pmatrix}F^t_\gamma&\sZ\end{pmatrix}\begin{pmatrix}0&F^t_{\gamma^{-1}}\\
-F_{\gamma^{-1}}&0\end{pmatrix}
\begin{pmatrix}F_\gamma\\\sZ^t\end{pmatrix}\\
&=-\sZ+\sZ^t
\end{align*}
Thus, $[\overline{\nabla},\overline{\nabla}]=0$ if and only if
$\sZ=\sZ^t$. In other words, the connection $\overline{\nabla}$
defines a holomorphic structure on $\mathcal{E}$ if and only if
$\sZ$ is symmetric. We note that a holomorphic structure on $\E$
corresponds to a real 2-form which is of type (1,1), as we will see
in below. Also, we may define a structure which corresponds to a
real 2-form which is of type (2,0) or (0,2). Using the holomorphic
connection $\overline{\nabla}$, we define a structure of type (2,0)
or (0,2) as $[\overline{\nabla},\nabla]=0$ and it is easy to see
that $\sZ=\overline{\sZ}^t$ if $[\overline{\nabla},\nabla]=0$.


Associated to the holomorphic structure $\overline{\nabla}$ on
$\mathcal{E}$, we may recover the compatible complex structure
$\widehat{J}_\Omega$ on $\T_\theta$ which is inherited from that of
$\T$. First note that the derivations are a noncommutative
generalization of derivatives $\frac{\partial}{\partial x}$. On the
other hand, the complex structure $\widehat{J}_\Omega$ in Section 2
is represented by using differential forms. Thus , in order to get a
matrix form for $\widehat{J}_\Omega$ on $\T_\theta$, we need to take
the dual period-like matrix $\mho^*$ and the relation between $\mho$
and $\mho^*$ is given as follows:
\[\begin{pmatrix}\mho^*\\\overline{\mho}^*\end{pmatrix}=
\begin{pmatrix}F_\gamma^t&\sZ\\F_\gamma^t&\overline{\sZ}\end{pmatrix}^{-t}=
\frac{1}{2i}\begin{pmatrix}-\text{Im }\sZ^{-t}&0\\0&\text{Im
}\sZ^{-t}\end{pmatrix}\begin{pmatrix}\overline{\sZ}^t&F_\gamma\\\sZ^t&F_\gamma\end{pmatrix}
\begin{pmatrix}F_{\gamma^{-1}}&0\\0&-F_{\gamma^{-1}}\end{pmatrix}\]
Thus, by a simple change of basis, we may set
$\mho^*=\begin{pmatrix}Z&F_\gamma\end{pmatrix}$ and as in Section 2,
we solve the matrix equation
\[\begin{pmatrix}Z&F_\gamma\end{pmatrix}\widehat{J}_\Omega=i\begin{pmatrix}Z&F_\gamma\end{pmatrix}\]
Using the identification $\rZ:V/L\cong L^*$, as in Section 2, we
have
\[\begin{pmatrix}Z&F_\gamma\cdot\rZ\end{pmatrix}\widehat{J}_\Omega
=i\begin{pmatrix}Z&F_\gamma\cdot\rZ\end{pmatrix}\] and hence
\begin{align}\label{ncstru}
\widehat{J}_\Omega=\begin{pmatrix}-(\text{Im }Z)^{-1}\text{Re }Z
&-(\text{Im
}Z)^{-1}\mathsf{F}_\gamma\\\mathsf{F}_{\gamma}^{-1}\left[\text{Im
}Z+\text{Re }Z(\text{Im }Z)^{-1}\text{Re
}Z\right]&\mathsf{F}_{\gamma}^{-1}\text{Re }Z(\text{Im
}Z)^{-1}\mathsf{F}_\gamma
\end{pmatrix},
\end{align}
where $\mathsf{F}_{\gamma}=F_{\gamma}\cdot\rZ$. Also, by the
relations (\ref{dcplex}) and (\ref{ncstru}), we get
\[\A=-(\text{Im }Z)^{-1}\text{Re }Z=
\mathsf{F}_{\gamma}^{-1}\text{Re }Z(\text{Im
}Z)^{-1}\mathsf{F}^t_\gamma.\] Furthermore, since $Z$ is symmetric,
we have the same relations as in (\ref{mirror}):
\[F_{\gamma}\cdot\rZ=\text{Im }Z \text{ \ \ and  \ \
}F_{\gamma}\cdot\iZ=\text{Re }Z.\]


Now by the same analysis given in Section 2, the matrix $\sZ$ is
symmetric if and only if the graph of the linear map
$f_{\gamma^{-1}}:V/L\lto L$ is an $\omega$-Lagrangian subspace of
$V$. Equivalently, the graph is a Lagrangian subspace of $V$ if and
only if the connection $\overline{\nabla}$ defines a compatible
connection on $\mathcal{E}$ on the noncommutative complex torus
$(\T_\theta,\widehat{J}_\Omega)$. Also, we see that the
antisymmetric bilinear form $E_{\gamma^{-1}}$, defined above,
satisfies the relation
\begin{equation}\label{real11}
E_{\gamma^{-1}}(\widehat{J}_\Omega v,\widehat{J}_\Omega
w)=E_{\gamma^{-1}}(v,w), \ \ v,w\in W.
\end{equation} From the relation (\ref{real11}), we find that
the graph $L_{\gamma^{-1}}=\{f_{\gamma^{-1}}(v)+v\mid v\in V/L\}$ of
the linear isomorphism $f_{\gamma^{-1}}:V/L\lto L$ corresponds to
the real 2-form $E_{\gamma^{-1}}$ of type $(1,1)$ and this implies
that the linear Lagrangian subspace $L_{\gamma^{-1}}$ of $V$ is
associated to a holomorphic structure on $\T_\theta$ via mirror
duality. However, $L_\gamma\cap\Gamma$ is not isomorphic to
$\mathbb{Z}^g$. In other words, the image
$\overline{L_{\gamma^{-1}}+\Gamma}$ of $L_{\gamma^{-1}}$ under the
covering map $V\lto V/\Gamma$ is not compact and is isomorphic to
$\mathbb{R}^g$ in $\mathbb{T}^d$.

Let us consider the case when the curvature $\gamma^{-1}$ is of the
most general form. Since $\gamma^{-1}$ is skew-symmetric, there is
an orthogonal matrix $O$ such that
\[O\gamma^{-1}O^t=\begin{pmatrix}0&\Delta_{\gamma^{-1}}\\-\Delta_{\gamma^{-1}}&0\end{pmatrix}\]
where $\Delta_{\gamma^{-1}}$ is a real $g\times g$ diagonal matrix.
Let
$\mho_O=\begin{pmatrix}\Delta_{\gamma^{-1}}^{-1}&\sZ\end{pmatrix}\cdot
O$ and let $\overline{\nabla}=\mho_O\nabla$. Then we have
\begin{align*}
[\overline{\nabla},\overline{\nabla}]&=[\mho_O\nabla,\mho_O\nabla]\\
&=\mho_O\gamma^{-1}\mho_O^t\\
&=\begin{pmatrix}\Delta_{\gamma^{-1}}^{-1}&\sZ\end{pmatrix}\cdot
O\gamma^{-1}O^t\begin{pmatrix}\Delta_{\gamma^{-1}}^{-1}\\\sZ^t\end{pmatrix}\\
&=\begin{pmatrix}\Delta_{\gamma^{-1}}^{-1}&\sZ\end{pmatrix}
\begin{pmatrix}0&\Delta_{\gamma^{-1}}\\-\Delta_{\gamma^{-1}}&0\end{pmatrix}
\begin{pmatrix}\Delta_{\gamma^{-1}}^{-1}\\\sZ^t\end{pmatrix}.
\end{align*}
Thus $\overline{\nabla}$ defines a holomorphic structure on
$\mathcal{E}$ if and only if $\sZ=\sZ^t$. In this case, associated
to the matrix
$\begin{pmatrix}0&\Delta_{\gamma^{-1}}\\-\Delta_{\gamma^{-1}}&0\end{pmatrix}$,
we have the Lagrangian subspace $L_{\Delta}$ of $V$ which is defined
to be the graph of the linear map $\Delta_{\gamma^{-1}}:V/L\lto L$.
Thus the corresponding Lagrangian subspace of $V$ for $\gamma^{-1}$
is given by the rotation of $L_\Delta$ by the orthogonal
transformation $O$.


Finally, we shall find holomorphic vectors for the vector bundle
$\mathcal{E}$ over $\T_\theta$. Recall that the holomorphic vectors
are elements of $H^0(\mathcal{E},\overline{\nabla})$, the kernel of
the linear map $\overline{\nabla}:\mathcal{E}\lto\mathcal{E}\otimes
(W^{0,1})^*$. Using the Euclidean metric for $W\cong\mathbb{R}^d$,
we may assume that the curvature matrix $\gamma^{-1}$ is of the form
given in (\ref{simple}). Then a compatible holomorphic structure on
$\mathcal{E}$ is specified by
$\overline{\nabla}=\mho_{\gamma}\nabla$, where
$\mho_\gamma=\begin{pmatrix}F_\gamma^t&\sZ\end{pmatrix}$ as given
above. By the definition of $\nabla$ given (\ref{connection}) we
have, for $\phi\in\mathcal{E}$ and
$\mathbf{s}=\begin{pmatrix}s_1&\cdots&s_g\end{pmatrix}^t\in\mathbb{R}^g$,
\begin{align*}
\begin{pmatrix}\overline{\nabla}_1\phi(\mathbf{s})\\\cdot\\\cdot\\\overline{\nabla}_g\phi(\mathbf{s})
\end{pmatrix}
&=\begin{pmatrix}F_\gamma^t&\sZ\end{pmatrix}
\begin{pmatrix}\nabla_1\phi(\mathbf{s})\\\cdot\\\cdot\\\cdot\\\cdot\\\nabla_d\phi(\mathbf{s})
\end{pmatrix}
=\begin{pmatrix}F_\gamma^t&\sZ\end{pmatrix}T^{-t}\begin{pmatrix}-(\frac{\partial\phi}{\partial
\mathbf{s}})^t\\2\pi i\mathbf{s}\phi(\mathbf{s})\end{pmatrix},
\end{align*}
where $\frac{\partial\phi}{\partial
\mathbf{s}}=\begin{pmatrix}\frac{\partial\phi}{\partial
s_1}&\cdots&\frac{\partial\phi}{\partial s_g}\end{pmatrix}$. Thus if
$\overline{\nabla}\phi(\mathbf{s})=0$, then we get a system of the
first order linear partial differential equations:
\begin{equation}\label{pde}
2\pi i\left(\sZ
T^{-t}_{[22]}+F_\gamma^tT^{-t}_{[12]}\right)\cdot\mathbf{s}-\left(
\sZ
T^{-t}_{[21]}+F_\gamma^tT^{-t}_{[11]}\right)\cdot\left(\frac{\partial\phi}{\partial
\mathbf{s}}\right)^t=0,
\end{equation}
where $T^{-t}_{[ij]}$ denotes the $g\times g$ matrix which
constitutes the $(i,j)$-block of $T^{-t}$. Then the solution for the
system (\ref{pde}) is
\[\phi(\mathbf{s})=\exp\left[\pi i\mathbf{s}^t\cdot\left(\sZ T^{-t}_{[21]}+F_\gamma^tT^{-t}_{[11]}\right)^{-1}
\left(\sZ T^{-t}_{[22]}+F_\gamma^tT^{-t}_{[12]}\right)\cdot
\mathbf{s}\right].\] Note that the set $\mathrm{M}_T$ of all
embeddings $T$ satisfying the relation (\ref{embedding}) is the
moduli space of finitely generated projective modules over
$\T_\theta$ equipped with a constant curvature connection $\nabla$
such that $[\nabla,\nabla]=2\pi i\gamma^{-1}$ and such a connection
is defined in terms of $T^{-t}$. Equivalently, the space
$\mathrm{M}_T$ is in one-to-one correspondence with the moduli space
of constant curvature connections on $\mathcal{E}$ whose curvature
is $2\pi i\gamma^{-1}$. Furthermore, if $\nabla_0$ is a connection
on $\mathcal{E}$ such that $[\nabla_0,\nabla_0]=2\pi i\gamma^{-1}$,
then all other connections satisfying the curvature condition are
given in the form $\nabla=\nabla_0+\mathbf{r}$, where
$\mathbf{r}\in\mathbb{R}^d$. Now we shall define a connection
$\nabla_0$ using the specific embedding
$T=\begin{pmatrix}F_\gamma^t&0\\0&1\end{pmatrix}$ or
$T^{-t}=\begin{pmatrix}F^{-t}_\gamma&0\\0&1\end{pmatrix}$. Then the
corresponding holomorphic structure is given by
$\overline{\nabla}_0=\mho_\gamma\nabla_0$ and the holomorphic vector
is
\[\phi(\mathbf{s})=\exp\left[\pi i\mathbf{s}^t\cdot
\sZ\cdot\mathbf{s}\right].\] Note that such a holomorphic vector
exists only if the complex matrix $\sZ$ is symmetric and thus if the
bundle does not admit a holomorphic structure than the holomorphic
vector does not exist. In general, for the connection
$\nabla=\nabla_0+\mathbf{r}$, the holomorphic connection is
analogously defined and the holomorphic vector is computed to be
\[\phi(\mathbf{s})=\exp\left[\pi i\mathbf{s}^t\cdot
\sZ\cdot\mathbf{s}+2\pi
i\mathbf{s}\cdot\mho_\gamma\cdot\mathbf{r}\right].\] Thus the
solution $\phi$ is in the Schwarz space $\mathcal{S}(\mathbb{R}^g)$
only when $\text{Im }\sZ>0$.

\section{Summary and Prospects}
In this paper, we have studied mirror duality on abelian varieties,
which has been well established. By an explicit study of the known
results, we could find an exact mirror correspondence between
complexfied symplectic form and the complex structure for the mirror
dual torus. Also, we have  reinterpreted the Riemann conditions in
terms of Lagrangian submanifolds and  we were able to find a
symplectic condition for non-holomorphic bundle over the mirror dual
torus. We found in Section 4 that all the above mentioned results
are naturally generalized to the case of holomorphic noncommutative
complex torus. The result for non-holomorphic bundle shed some
lights to the study of non-holomorphic noncommutative complex torus
and we will address to this problem later \cite{KK1}. Associated to
this problem, We described the mirror structure for generalized
complex torus with canonical complex structure in Section 2. First
we have rephrased the mirror duality on abelian varieties in terms
of generalized complex structures in the case when the given B-field
is of type (1,1). Also, we discussed the case when B-field is a
general type. This might be a first step for us to go for studying
Kontsevich's homological mirror symmetry for abelian varieties
equipped with a general type of B-field as was indicated in
\cite{Kap}. We will study this problem too later (\cite{KK2}) with
the categorical approach using the Lie algebroid structure as
studied in \cite{Bl}.


\begin{center}
{\large \bf Acknowledgments} \\
\end{center}
Hoil Kim is supported by KRF-2004.

\end{document}